\documentclass[aps,pre,twocolumn,superscriptaddress,nofootinbib]{revtex4-2}
\usepackage{amsmath,amssymb,bm}
\usepackage{graphicx}
\usepackage{physics}
\usepackage{hyperref}
\usepackage{microtype}

\newcommand{\tS}{\mbox{\tiny S}}
\newcommand{\tB}{\mbox{\tiny B}}
\newcommand{\loc}{\mathrm{loc}}
\newcommand{\MF}{\mathrm{MF}}
\newcommand{\ee}{\mathrm{e}}

\newcommand{\wti}{\widetilde}

\newcommand{\Eq}[1]{Eq.\,(\ref{#1})}
\newcommand{\Eqs}[1]{Eqs.\,(\ref{#1})}
\newcommand{\Fig}[1]{Fig.\,\ref{#1}}
\newcommand{\bsube}{\begin{subequations}}
\newcommand{\esube}{\end{subequations}}
\newcommand{\dg}{\dagger}

\begin{document}

\title{Temperature--Hamiltonian Ambiguity in Strong-Coupling Quantum Equilibrium}

\author{Jinshuang Jin}
\email{jsjin@hznu.edu.cn}
\affiliation{School of Physics, Hangzhou Normal University, Hangzhou, Zhejiang 311121, China}

\date{\today}

\begin{abstract}
Strong coupling generally drives a subsystem away from the Gibbs state of its
bare Hamiltonian at the bath temperature. We formulate a
temperature--Hamiltonian ambiguity: the reduced equilibrium state fixes the
dimensionless Gibbs generator, but does not by itself uniquely determine the
local Hamiltonian and temperature entering a Gibbs representation. We propose
an operational resolution in which the local Hamiltonian is identified from
physical information independent of the stationary Gibbs parametrization.
Whenever the exact reduced state is compatible with the Gibbs family generated
by that independently identified Hamiltonian, it then determines a unique
local temperature. An exactly solvable two-level-system--oscillator model makes the
ambiguity explicit through state-equivalent descriptions based on temperature
renormalization relative to a specified local Hamiltonian and
Hamiltonian-of-mean-force level renormalization at the bath temperature. An
exact bosonic model provides the complementary case in which retarded dynamics
independently fixes a renormalized local excitation scale and thereby specifies
the local Hamiltonian before the stationary populations determine the
temperature. These results distinguish state representation from
thermodynamic assignment and provide an operational framework for local
thermodynamics beyond weak coupling.
\end{abstract}

\maketitle

Strong coupling generally drives an open-system equilibrium state away from the
Gibbs state of its bare Hamiltonian and makes it explicitly dependent on the
environment and coupling~\cite{Tal20041002,Tru22012301}. An exact representation
is provided by the Hamiltonian of mean force (HMF),
\begin{equation}
 H_{\MF}
 =
 -\frac{1}{\beta}
 \ln\!\left[
 \frac{\Tr_{\tB}\ee^{-\beta H_{\rm tot}}}{Z_{\tB}^{0}}
 \right],
 \qquad
 \rho_{\tS}=\frac{\ee^{-\beta H_{\MF}}}{Z_{\MF}},
 \label{eq:HMF}
\end{equation}
which underlies strong-coupling thermodynamics~\cite{Tal20041002,Tru22012301,Tru22042209},
while its temperature dependence and associated assignment freedom complicate
a direct local interpretation~\cite{Str20050101}. Strong coupling can also
renormalize local spectra and effective temperatures, as found in
reaction-coordinate, Gaussian, exact dynamical, Brownian-motion, and finite-bath
approaches~\cite{Ant23020307,Hsi21065001,Hua22023141,Yao24085114,Du25024139}.

Related nonuniqueness has appeared in other settings: effective-temperature
assignments can depend on the parametrization of a strongly coupled Gaussian
state~\cite{Hsi21085004}; entanglement-temperature mappings can admit equivalent
Hamiltonian/temperature interpretations~\cite{Mog22030335}; and common energy--temperature
rescaling is already present in the Boltzmann distribution~\cite{Bra1900444}.
Recent nonequilibrium work likewise emphasizes that subsystem temperature need
not follow from the reduced state alone~\cite{Fag260708655}. Here we formulate
the corresponding equilibrium identifiability problem. A full-rank
$\rho_{\tS}$ fixes $-\ln\rho_{\tS}$ up to an additive scalar, but not its
separation into an inverse temperature and an energy operator. This
multiplicative nonuniqueness, distinct from the additive HMF freedom
~\cite{Str20050101}, is the \emph{temperature--Hamiltonian ambiguity}.

We resolve it operationally and conditionally: identify a physically
motivated local Hamiltonian using information independent of the stationary
Gibbs parametrization, and then test whether the exact reduced state belongs to
its Gibbs family. If it does, the state fixes a unique local temperature
relative to that Hamiltonian; if it does not, no single scalar local
temperature is assigned. The identification of the local Hamiltonian itself is
therefore model- or measurement-specific rather than universal. A
quantum-nondemolitional two-level-system--oscillator model makes the ambiguity
explicit through two state-equivalent Gibbs representations, whereas an exact
bosonic model~\cite{Hua22023141} shows how retarded dynamics can independently
calibrate a renormalized local excitation scale before stationary populations
determine the temperature.

\paragraph*{General principle.---}
Let $\rho_{\tS}=\Tr_{\tB}\rho_{\rm tot}$ be an exact stationary reduced state
of a composite system in thermal equilibrium,
$\rho_{\rm tot}\propto\ee^{-\beta H_{\rm tot}}$, with $\beta=1/(k_{\tB}T)$.
For a full-rank state, one may formally write
\begin{equation}
 \rho_{\tS}=
 \frac{\ee^{-K_{\tS}}}{\Tr_{\tS}\ee^{-K_{\tS}}},
 \label{eq:modular}
\end{equation}
where $K_{\tS}$ is determined by $\rho_{\tS}$ up to an additive constant (a multiple of the identity).
A Gibbs interpretation amounts to the factorization
\begin{equation}
 K_{\tS}=\beta^{*}H_{\tS}^{\loc}+cI, 
 \label{eq:factorization}
\end{equation}
where $\beta^*=1/(k_{\tB}T^*)$ and $c$ is an arbitrary scalar.
The reduced state alone therefore does not uniquely separate a temperature scale from a local
Hamiltonian.
Our prescription is to identify $H_{\tS}^{\loc}$ by an independent physical
criterion and then test Gibbs compatibility. A single local temperature exists
relative to this Hamiltonian only when there are a scalar $\beta^*$ and an
additive constant $c$ such that Eq.~(\ref{eq:factorization}) holds. Equivalently,
the exact state must belong to the Gibbs family generated by the independently
identified $H_{\tS}^{\loc}$. When this condition is satisfied,
\begin{equation}
 \rho_{\tS}=
 \frac{\ee^{-\beta^{*}H_{\tS}^{\loc}}}{Z_{\tS}},
 \qquad
 Z_{\tS}=\Tr_{\tS}\ee^{-\beta^{*}H_{\tS}^{\loc}},
 \label{eq:localGibbs}
\end{equation}
defines $\beta^*$, and hence the local effective temperature. If the
compatibility condition fails, no single scalar temperature can reproduce the
exact state relative to that Hamiltonian.
Hamiltonian and temperature renormalizations are thereby treated as conceptually distinct,
\begin{equation}
 H_{\tS}\rightarrow H_{\tS}^{\loc},
 \qquad
 T\rightarrow T^{*}.
 \label{eq:twoRenorm}
\end{equation}
After $H_{\tS}^{\loc}$
 has been independently specified, it is regarded
 as independent of $\beta^{*}$ in the following canonical construction.

The corresponding local internal energy and entropy are given by
\bsube\label{US0}
\begin{align}
 U_{\tS}&=\Tr_{\tS}(H_{\tS}^{\loc}\rho_{\tS}),
 \label{eq:Uloc}\\
 S_{\tS}&=-k_{\tB}\Tr_{\tS}(\rho_{\tS}\ln\rho_{\tS}).
 \label{eq:Sloc}
\end{align}
\esube
Once $H_{\tS}^{\loc}$ has been independently fixed, 
the Gibbs relation in Eq.~(\ref{eq:localGibbs}) serves as our primary definition of $\beta^{*}$, 
rather than as a thermodynamic derivative.
For the canonical family generated by this fixed Hamiltonian--which contains 
  no explicit $\beta^{*}$ 
 dependence--the standard identities then follow:
 \bsube\label{TC}
\begin{align}
 T^{*}
 &=\left.\frac{\partial U_{\tS}}{\partial S_{\tS}}
 \right|_{H_{\tS}^{\loc}},
 \label{Teff0}\\
 C_{\tS}^{\loc}
 &\equiv
 \left.\frac{\partial U_{\tS}}{\partial T^{*}}
 \right|_{H_{\tS}^{\loc}}
 =
 \frac{\langle(H_{\tS}^{\loc}-\langle H_{\tS}^{\loc}\rangle)^2\rangle}
 {k_{\tB}(T^{*})^2}\ge0 .
 \label{heatC0}
\end{align}
\esube
Likewise, the Gibbs form ensures~\cite{Ste10}
\bsube\label{US1}
\begin{align}
 U_{\tS}&=
 - \partial_{\beta^\ast}\ln Z_{\tS},
 \label{eq:Uloc1}\\
 S_{\tS}&=
 k_{\tB}\beta^{\ast2}
  \partial_{\beta^\ast}F_{\tS},
 \label{eq:Sloc1}
\end{align}
where
\begin{equation}
 F_{\tS}=U_{\tS}-T^{*}S_{\tS}
       =-\frac{1}{\beta^{*}}\ln Z_{\tS}.
 \label{eq:Floc1}
\end{equation}
\esube
The fixed-Hamiltonian qualifier in \Eq{TC} is essential.
Along a physical coupling trajectory, $H_{\tS}^{\loc}$ may itself vary, so
$dU_{\tS}$ contains both population and spectral contributions. Such a total
derivative should not be confused with the canonical derivatives above.

This construction does not challenge the exact HMF identity in
Eq.~(\ref{eq:HMF}). Rather, it separates an exact representation of the
reduced state and its associated excess thermodynamics from local
thermodynamics defined with respect to an independently identified energy
operator. Indeed, because $H_{\MF}$ generally depends on $\beta$,
 $-\partial_{\beta}\ln Z_{\MF}
 =\left\langle H_{\MF}+\beta\,\partial_{\beta}H_{\MF}\right\rangle$,
which need not coincide with $\Tr_{\tS}(H_{\tS}^{\loc}\rho_{\tS})$
or $\Tr_{\tS}(H_{\MF}\rho_{\tS})$. Likewise, in general,
$S_{\tS}\neq k_{\tB}\beta^{2}\partial_{\beta}F_{\MF}$.
The central prescription is therefore
\begin{equation}
 \boxed{
 \begin{gathered}
 H_{\tS}^{\loc}\ {\rm identified\ independently}
  +  \rho_{\tS}\ {\rm exact}\\[-1mm]
  +\,  {\rm Gibbs\ compatibility}
 \end{gathered}
 \Longrightarrow T^{*}.}
 \label{eq:principle}
\end{equation}


\paragraph*{Exactly solvable TLS example.---}
Consider a two-level system (TLS) coupled quantum-nondemolitionally (QND)
to a harmonic oscillator \cite{Cam09392002},
\begin{equation}
 H_{\rm tot}
 =\frac{\varepsilon}{2}\sigma_z
 +\omega\left(b^{\dagger}b+\frac12\right)
 +\chi\sigma_z\left(b^{\dagger}b+\frac12\right),
 \label{eq:TLSHam}
\end{equation}
with $|\chi|<\omega$. 
For $\sigma=\pm1$, the oscillator frequency conditioned on the TLS state is
$\omega_{\sigma}=\omega+\sigma\chi$.
In equilibrium, the exact partition functions are
$Z_{\rm tot}={\rm Tr}e^{-\beta H_{\rm tot}}\equiv\sum_\sigma z_\sigma$, with
\begin{equation}
 z_{\sigma}
 =
 \frac{\ee^{-\sigma\beta\varepsilon/2}}
 {2\sinh[\beta(\omega+\sigma\chi)/2]} .
 \label{eq:zsigma}
\end{equation}
The reduced TLS state is therefore
\begin{equation}
 \rho_{\tS} 
 =p_{+}\ket{+}\bra{+}+p_{-}\ket{-}\bra{-},
 \qquad
 p_{\sigma}=\frac{z_{\sigma}}{z_{+}+z_{-}} .
 \label{eq:TLSrho}
\end{equation}
Writing the same state as
$\rho_{\tS}\propto\ee^{-\beta^{*}H_{\tS}^{\loc}}$ with
$H_{\tS}^{\loc}=(\varepsilon_{\loc}/2)\sigma_z$ gives
\begin{equation}
 \beta^{*}\varepsilon_{\loc}=\ln\frac{p_{-}}{p_{+}},
 \label{eq:TLSambiguity}
\end{equation}
which is the elementary form of the temperature--Hamiltonian ambiguity.

Because the QND coupling preserves the bare TLS eigenbasis, this model provides
a particularly transparent setting in which to expose the ambiguity. We
choose the independently specified bare TLS Hamiltonian
$H_{\tS}^{\loc}=H_{\tS}=(\varepsilon/2)\sigma_z$ as the local energy
reference and ask which temperature reproduces the exact reduced populations.
This is an operational choice, not a consequence of the reduced state itself:
fixing the bath temperature instead leads to the equivalent HMF level
renormalization discussed later. With this local energy reference,
\begin{equation}
 \beta^{*}
 =
 \beta\left(1-\frac{\chi}{\varepsilon}\right)
 +\frac{1}{\varepsilon}
 \ln\!\left[
 \frac{\ee^{\beta(\omega+\chi)}-1}
      {\ee^{\beta(\omega-\chi)}-1}
 \right].
 \label{eq:betaStarTLS}
\end{equation}
The corresponding temperature follows directly from the Gibbs population
ratio [\Eq{eq:TLSambiguity}],
\begin{equation}
 T^{*}
 =
 \frac{\varepsilon}
 {k_{\tB}\ln(p_{-}/p_{+})},
 \label{eq:TStarTLS}
\end{equation}
and, for the canonical family at fixed $\varepsilon$, it also satisfies
$T^{*}=(\partial U_{\tS}/\partial S_{\tS})_{\varepsilon}$.
The corresponding local partition function is
\begin{equation}\label{ZS0}
Z_{\tS}
=2\cosh\!\left(\frac{\beta^\ast\varepsilon}{2}\right)
=\frac{Z_{\rm tot}}{\sqrt{z_+z_-}}.
\end{equation}
The associated canonical local heat capacity is
\begin{equation}
 C_{\tS}^{\loc}
 =
 k_{\tB} p_{+}p_{-}
 \left[\ln\frac{p_{-}}{p_{+}}\right]^2
 \ge 0 .
 \label{eq:ClocTLS}
\end{equation}

\begin{figure}[t]
 \centering
\centerline{\includegraphics*[width=1.0\columnwidth,angle=0]{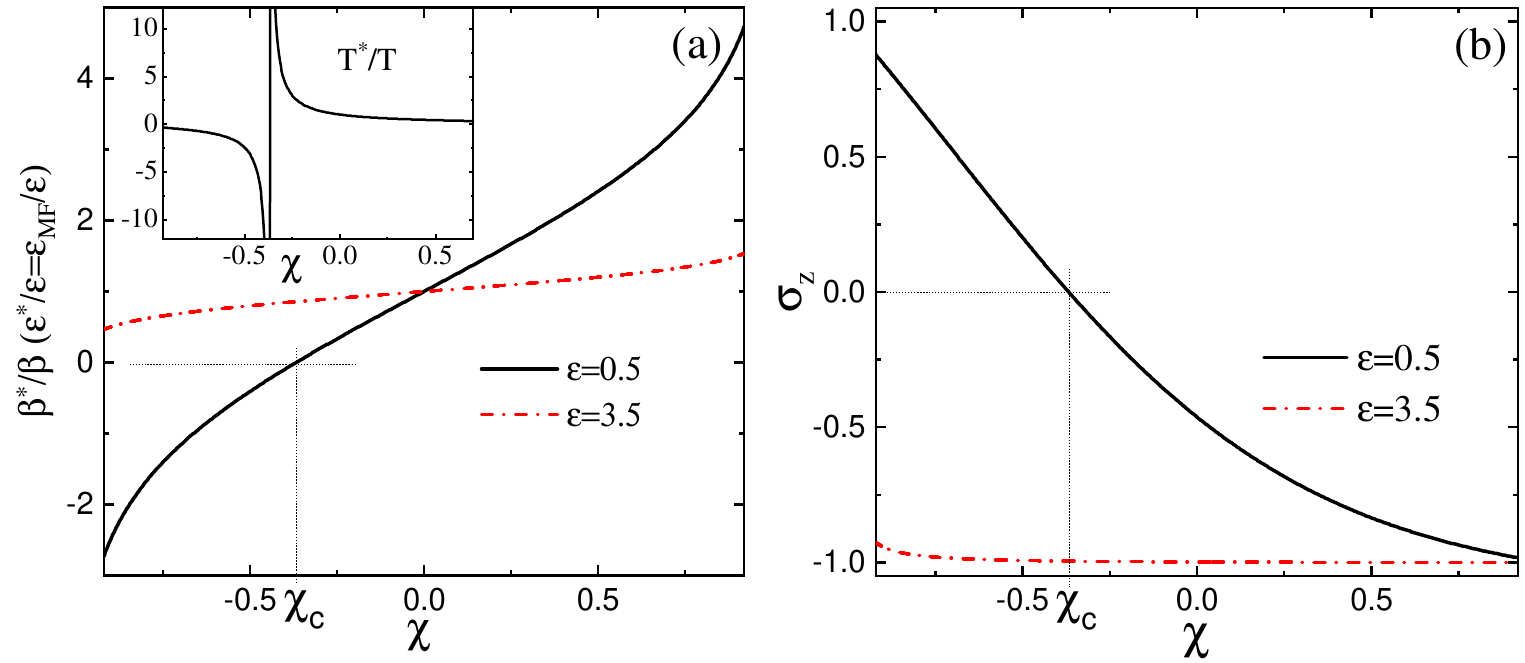}}
 \caption{
Temperature--Hamiltonian ambiguity in the TLS for $k_{\tB}T=\omega=1$.
The curves correspond to $\varepsilon=0.5$ and $3.5$. For $\varepsilon=0.5$,
the nonsingular ratio $\beta^{*}/\beta$ crosses zero at $\chi_c\simeq-0.2273$,
where the reduced TLS is maximally mixed; the pole in $T^{*}/T$ is therefore
the regular infinite-temperature point. Panel (b) shows the accompanying
population inversion.
 }
 \label{fig1-TLS}
\end{figure}

Figure~\ref{fig1-TLS}(a) shows the resulting coupling dependence.
For the parameter regime considered here, the TLS also exhibits an
instructive zero crossing of $\beta^{*}$.
Define the critical coupling $\chi_c$ by
\begin{equation}
 \beta^{*}(\chi_c)=0
 \Longleftrightarrow
 z_{-}(\chi_c)=z_{+}(\chi_c),
 \label{eq:chicdef}
\end{equation}
At this point,
\begin{equation*}
 p_{+}=p_{-}=\frac12,\qquad
 \rho_{\tS}=\frac{I}{2},
 \end{equation*}
 where
$ U_{\tS}=0$, $ C_{\tS}^{\loc}=0$, and $S_{\tS}=k_{\tB}\ln2$.
Thus the divergence
$T^{*}=1/(k_{\tB}\beta^{*})\rightarrow\pm\infty$
does not signal a singular reduced state or a phase transition; it is the
regular infinite-temperature point of the bounded TLS spectrum.
For $\varepsilon>0$,
\begin{equation}
 \beta^{*}<0 \Longleftrightarrow p_{+}>p_{-},
 \label{eq:negativeT}
\end{equation}
so the negative-$T^{*}$ branch is precisely a population-inverted reduced state relative to
the fixed bare TLS Hamiltonian, as shown in \Fig{fig1-TLS} (b).
The bath temperature remains positive. This negative-temperature branch is
specific to the bounded TLS spectrum and is not a generic consequence of
strong coupling or of the temperature--Hamiltonian ambiguity itself.

The same state may instead be represented at the bath temperature by
$H_{\tS}^{*}(\beta)=[\varepsilon^{*}(\beta)/2]\sigma_z$, i.e., $\rho_{\tS}\propto\ee^{-\beta H^{*}_{\tS}(\beta)}$,
with
\begin{equation}
 \varepsilon^{*}(\beta)
 =
 \varepsilon-\chi
 +\frac{1}{\beta}
 \ln\!\left[
 \frac{\ee^{\beta(\omega+\chi)}-1}
      {\ee^{\beta(\omega-\chi)}-1}
 \right].
 \label{eq:epsstar}
\end{equation}
For the HMF \Eq{eq:HMF} (see Supplemental Material~\cite{SM}, Sec.~III, for the explicit HMF thermodynamics),
\begin{equation}
 H_{\MF}=C(\beta,\chi)I+\frac{\varepsilon^{*}(\beta)}{2}\sigma_z ,
 \label{eq:HMFTLS}
\end{equation}
where $C(\beta,\chi) = -\frac{1}{2\beta}\ln
 [z_+z_-/(Z_{\rm B}^0)^2]$. Consequently,
\begin{equation}
 \frac{\ee^{-\beta^{*}H_{\tS}}}{Z_{\tS}}
 =
 \frac{\ee^{-\beta H_{\tS}^{*}(\beta)}}{Z_{\tS}^{*}}
 =
 \frac{\ee^{-\beta H_{\MF}}}{Z_{\MF}}
 =
 \rho_{\tS}.
 \label{eq:stateequiv}
\end{equation}
These representations describe the same state but assign different
thermodynamic parameters.
In particular,
\begin{equation}
 \frac{\beta^{*}}{\beta}
 =
 \frac{\varepsilon^{*}}{\varepsilon}
 =
 \frac{\varepsilon_{\MF}}{\varepsilon},
 \qquad
 \frac{T^{*}}{T}
 =
 \frac{\varepsilon}{\varepsilon_{\MF}},
 \label{eq:equivratio}
\end{equation}
so the same crossing is described either as a temperature inversion at fixed $\varepsilon$
or as a level inversion at fixed bath temperature.

\paragraph*{Open bosonic mode: resolving the ambiguity dynamically.---}
The TLS example exposes the ambiguity using the bare TLS Hamiltonian as an
explicitly chosen local energy reference. It does not, however, address the
complementary situation in which independent dynamical information reveals a
renormalized local spectral scale.
An exactly solvable bosonic model provides an instructive realization of
our prescription when the interaction genuinely renormalizes the local
dynamics. We use this established solution not to rederive strong-coupling
renormalization, but to demonstrate the operational principle in
Eq.~(\ref{eq:principle}).

We therefore consider a single-mode bosonic system coupled to a continuum,
\begin{equation}
 H_{\rm tot}
 =
 \omega_{\tS}a^{\dagger}a
 +\sum_k\omega_k b_k^{\dagger}b_k
 +\sum_k\left(
 V_k a^{\dagger}b_k+V_k^{*}b_k^{\dagger}a
 \right).
 \label{eq:Fano}
\end{equation}
This Fano--Anderson model and its exact non-Markovian dynamics are well
established~\cite{Hua22023141,Tu08235311,Jin10083013,Zha12170402}.
Here the retarded dynamics supplies information independent of the stationary
Gibbs parametrization. For this number-conserving model it determines a
natural renormalized local excitation scale, which in turn specifies the
local Hamiltonian. In particular, the exact dynamics gives
~\cite{Hua22023141,Tu08235311,Jin10083013,Zha12170402}
\begin{equation}
 H_{\tS}^{r}(t)=\omega_{\tS}^{r}(t)a^{\dagger}a,
 \qquad
 \omega_{\tS}^{r}(t)
 =
 -\Im\frac{\dot u(t,t_0)}{u(t,t_0)} ,
 \label{eq:omegaR}
\end{equation}
where $u(t,t_0)$ is the exact retarded propagator; the corresponding Dyson equation and spectral representation are summarized in Supplemental Material~\cite{SM}, Sec.~II.
In the stationary continuum regime considered below, whenever the asymptotic
limit exists, we use the renormalized excitation frequency extracted from the
retarded propagator to specify the local Hamiltonian,
\begin{equation}
 H_{\tS}^{\loc}
 =\wti{\omega}_{\tS}a^{\dagger}a,
 \qquad
 \wti{\omega}_{\tS}\equiv
 \lim_{t\rightarrow\infty}\omega_{\tS}^{r}(t).
 \label{eq:HlocBoson}
\end{equation}
This identification is model-specific rather than a universal equivalence
between a time-local dynamical Hamiltonian and a thermodynamic Hamiltonian.
Its role here is operational: the retarded dynamics identifies
$H_{\tS}^{\loc}$ independently of the stationary populations. For the
numerical results below, the same stationary excitation scale is evaluated
from the equivalent self-energy resonance condition given in Supplemental
Material~\cite{SM}, Sec.~II.

In the continuum regime, the exact stationary reduced state has the
geometric form~\cite{Hua22023141,Hua20165116}
\begin{equation}
 \rho_{\tS}
 =
 \sum_{n=0}^{\infty}
 \frac{\bar n_{\tS}^{\,n}}
 {(1+\bar n_{\tS})^{n+1}}
 \ket{n}\bra{n},
 \label{eq:rhoBoson}
\end{equation}
Once this local energy scale is fixed independently, the
temperature--Hamiltonian ambiguity is removed. Writing
Eq.~(\ref{eq:rhoBoson}) as
$\rho_{\tS}\propto\ee^{-\beta^{*}\wti\omega_{\tS}a^\dg a}$ gives 
\begin{equation}
\label{betaeff1}
\beta^\ast\wti\omega_{\tS}=
\ln\frac{1+\bar n_{\tS}}{\bar n_{\tS}}.
\end{equation}
In the absence of an additional localized-mode contribution,
$\bar n_{\tS}={\rm Tr}_{\tS}(\rho_{\tS}a^\dg a)$ is exactly
~\cite{Hua22023141} (see Supplemental Material~\cite{SM}, Sec.~II, for the spectral representation and bound-state criterion),
\begin{equation}
 \bar n_{\tS}
 =
 \int_0^{\infty}\frac{d\omega}{2\pi}
 D(\omega)n_{\tB}(\omega),
 \label{eq:nbarBoson}
\end{equation}
where
$n_{\tB}(\omega)=[\ee^{\beta\omega}-1]^{-1}$ is the initial bosonic
occupation of the reservoir, and $D(\omega)$ is determined by the reservoir
spectral density
$J(\omega)=2\pi\sum_k|V_k|^2\delta(\omega-\omega_k)$.

\begin{figure}[t]
 \centering
\centerline{\includegraphics*[width=1.0\columnwidth,angle=0]{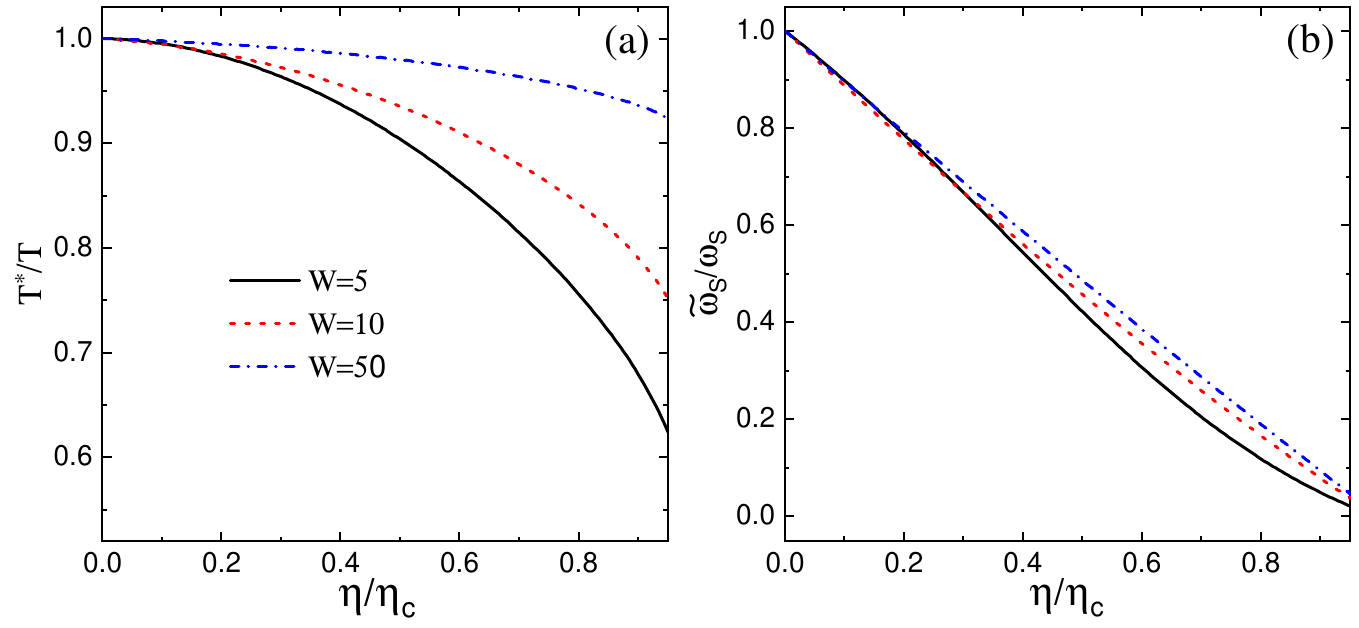}}
  \caption{
Independent Hamiltonian and temperature renormalizations for the
Lorentz--Drude reservoir with $k_{\tB}T=\omega_{\tS}=1$ and
$W/\omega_{\tS}=5,10,50$. The effective temperature $T^{*}$ is inferred from
the stationary occupation after the local excitation scale has been fixed,
whereas the renormalized frequency $\wti{\omega}_{\tS}$ is determined by the
retarded dynamics. For each bandwidth the coupling is normalized by the
bound-state threshold $\eta_c=2\omega_{\tS}/(\pi W)$; see Supplemental Material~\cite{SM}, Sec.~II, for the spectral representation.
 }
 \label{fig2-boson}
\end{figure}

With the independently determined $H_{\tS}^{\loc}$, Eqs.~(\ref{eq:Uloc})
and (\ref{eq:Sloc}) become
\bsube
\begin{align}
 U_{\tS}&=\wti{\omega}_{\tS}\bar n_{\tS},
 \label{eq:UBoson}\\
 S_{\tS}
 &=
 k_{\tB}(1+\bar n_{\tS})\ln(1+\bar n_{\tS})
 -k_{\tB}\bar n_{\tS}\ln\bar n_{\tS}.
 \label{eq:SBoson}
\end{align}
\esube
and therefore yield the local effective temperature
\begin{equation}
 T^{*} =
 \left.\frac{\partial U_{\tS}}{\partial S_{\tS}}\right|_{\wti{\omega}_{\tS}}
 =
 \frac{\wti{\omega}_{\tS}}
 {k_{\tB}\ln[(1+\bar n_{\tS})/\bar n_{\tS}]},
 \label{eq:TStarBoson}
\end{equation}
which is consistent with \Eq{betaeff1}.
At fixed $\wti{\omega}_{\tS}$,
\begin{equation}
 C_{\tS}^{\loc}
 =
 k_{\tB}(\beta^{*}\wti{\omega}_{\tS})^2
 \bar n_{\tS}(1+\bar n_{\tS})
 \ge0 .
 \label{eq:CBoson}
\end{equation}
The two pieces of information are therefore operationally distinct: the
retarded Green function calibrates the local excitation energy in this model,
whereas the stationary reduced density matrix determines the effective
temperature relative to that calibrated spectrum.
The effective temperature is thus not another name for the frequency
renormalization; it is the temperature that reproduces the exact stationary
populations relative to the independently determined local spectrum.
In the weak-coupling limit both descriptions continuously recover the
bare equilibrium result, $T^*\to T$ and
$\widetilde{\omega}_S\to\omega_S$.  At finite coupling, however, the
two renormalizations display distinct dependences on the reservoir
bandwidth, confirming that they encode different physical information,
as can be seen in \Fig{fig2-boson}.


\paragraph*{Local versus HMF thermodynamics.---}
The two examples realize the same operational structure in complementary
ways. For the TLS, the independently specified bare Hamiltonian is adopted as
$H_{\tS}^{\loc}$. For the bosonic model, the asymptotic retarded dynamics
determines a renormalized local spectral scale and thereby specifies
$H_{\tS}^{\loc}$. In both cases the effective temperature is inferred only
after the local Hamiltonian has been identified independently of the
stationary Gibbs parametrization.
The resulting local thermodynamic identities are therefore mutually
consistent and share the same fixed-Hamiltonian interpretation.

State equivalence does not imply equivalence of thermodynamic
interpretations [see Eq.~(\ref{eq:equivratio})].
The HMF remains an exact representation of the reduced equilibrium
state. Thermodynamic quantities generated from
$Z_{\MF}=Z_{\rm tot}/Z_{\tB}^{0}$, however, are defined relative to a
bare-bath ($Z_{\tB}^{0}$) reference and therefore need not coincide with local-state
quantities such as $\Tr_{\tS}(H_{\tS}^{\loc}\rho_{\tS})$.
Moreover, a temperature-dependent effective Hamiltonian generates
additional $\beta\partial_{\beta}H^\ast(\beta)$ terms in thermodynamic
derivatives. Once a physical local Hamiltonian is independently fixed, the
remaining population renormalization instead has a direct local
thermodynamic interpretation through $T^{*}$.

\begin{figure}[t]
 \centering
\centerline{\includegraphics*[width=1.02\columnwidth,angle=0]{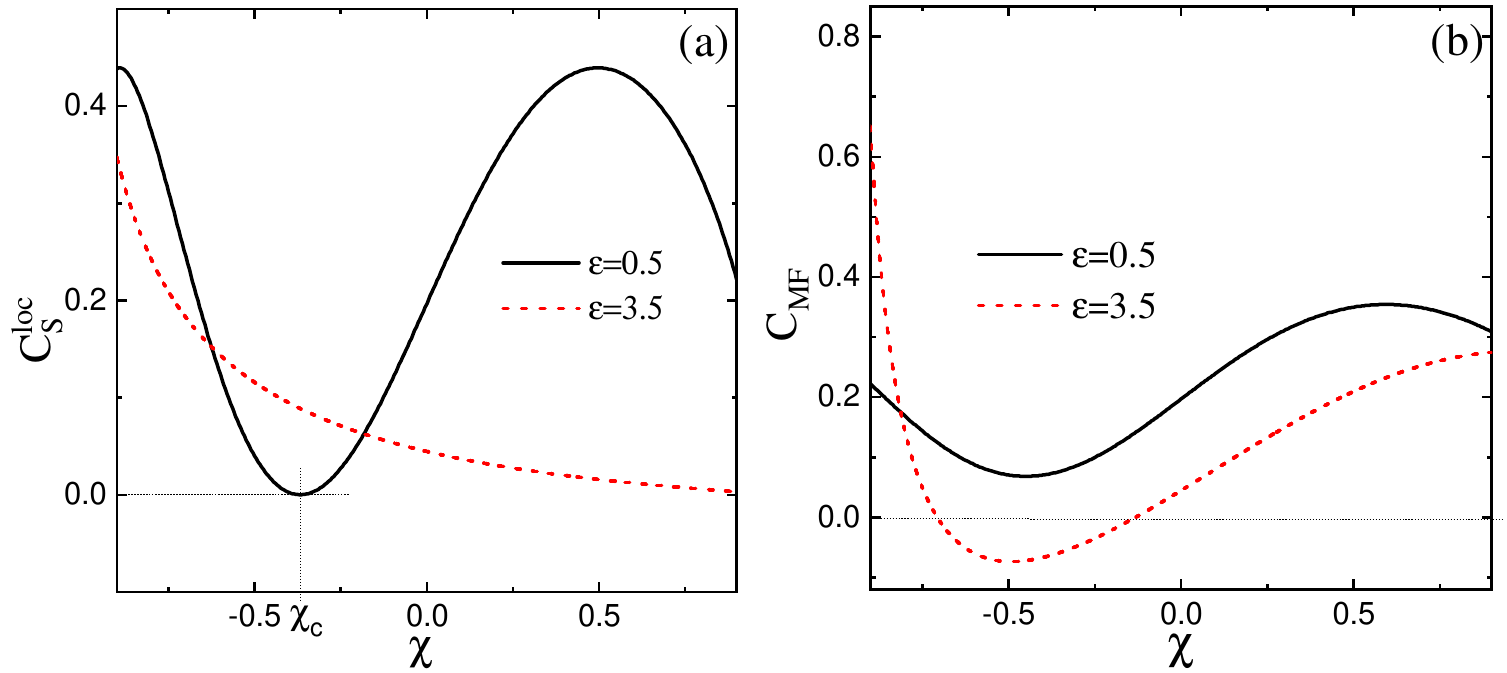}}
 \caption{Local versus HMF thermodynamic responses for $k_{\tB}T=\omega=1$ and
$\varepsilon=0.5,3.5$. The local heat capacity is a canonical response defined
by the independently identified local Hamiltonian and temperature, whereas
$C_{\MF}$ is an excess response of the interacting composite relative to the
bare bath.
 }
 \label{fig3}
\end{figure}

For the TLS one may compare (see Supplemental Material~\cite{SM}, Sec.~III, for the explicit HMF thermodynamics and heat-capacity derivation)
\bsube
\begin{align}
 U_{\MF}&=-\partial_{\beta}\ln Z_{\MF}=\left\langle H_{\MF}+\beta\,\partial_{\beta}H_{\MF}\right\rangle,
 \\
 C_{\MF}&=\frac{\partial U_{\rm MF}}{\partial T}=C_{\rm tot}-C_{\tB}^{0},
 \label{eq:CMF}
\end{align}
\esube
with the canonical local quantities in Eqs.~(\ref{eq:Uloc}) and (\ref{heatC0}).
As illustrated in Fig.~\ref{fig3}, a negative $C_{\MF}$, when it occurs,
should therefore be interpreted as a negative excess response relative to the
bare bath, not as a negative canonical heat capacity of the local TLS.
For the present two examples, \Eqs{eq:ClocTLS} and (\ref{eq:CBoson}) guarantee
$C_{\tS}^{\loc}\ge0$ in the local Gibbs description.

\paragraph*{Conclusion.---}
A reduced equilibrium state fixes a dimensionless Gibbs generator but does not
by itself uniquely determine the local Hamiltonian and temperature entering a
Gibbs representation. We resolve this temperature--Hamiltonian ambiguity
operationally by identifying the local Hamiltonian from physical information
independent of the stationary Gibbs parametrization and then testing Gibbs
compatibility. When the exact state belongs to the Gibbs family generated by
that Hamiltonian, it determines the corresponding local temperature; when it
does not, no single local temperature can be assigned relative to that local
energy operator. The TLS model
makes the ambiguity explicit: the same reduced state admits a
temperature-renormalized description relative to the specified bare
Hamiltonian and an equivalent mean-force Hamiltonian-renormalized description
at the bath temperature. The bosonic model provides the complementary case in
which retarded dynamics independently determines a renormalized local
excitation scale and thereby specifies the local Hamiltonian before the
stationary populations determine $T^{*}$. State-equivalent Gibbs
representations therefore need not imply equivalent thermodynamic
assignments; an independently identified local energy scale is the additional
physical input required to define local thermodynamics beyond weak coupling.

\paragraph*{Data availability.---}
The numerical data supporting the figures can be reproduced from the
equations and parameters given in the main text and Supplemental Material.
The numerical data and scripts are available from the corresponding author
upon reasonable request.

\begin{acknowledgments}
We acknowledge helpful discussions with Prof.\ Wei-Min Zhang and Prof.\ YiJing Yan.
OpenAI ChatGPT (GPT-5.6 Sol) was used to assist with manuscript organization,
language editing, algebraic checks, and numerical-code development. The author
is responsible for the derivations, numerical results, figures, references,
and scientific conclusions.
\end{acknowledgments}


\begin{thebibliography}{99}

\bibitem{Tal20041002}
P. Talkner and P. H\"anggi,
Rev. Mod. Phys. \textbf{92}, 041002 (2020).

\bibitem{Tru22012301}
A. S. Trushechkin, M. Merkli, J. D. Cresser, and J. Anders,
AVS Quantum Sci. \textbf{4}, 012301 (2022).

\bibitem{Tru22042209}
A. Trushechkin,
Phys. Rev. A \textbf{106}, 042209 (2022).

\bibitem{Str20050101}
P. Strasberg and M. Esposito,
Phys. Rev. E \textbf{101}, 050101(R) (2020).

\bibitem{Ant23020307}
N. Anto-Sztrikacs, A. Nazir, and D. Segal,
PRX Quantum \textbf{4}, 020307 (2023).

\bibitem{Hsi21065001}
J.-T. Hsiang and B.-L. Hu,
Phys. Rev. D \textbf{103}, 065001 (2021).

\bibitem{Hua22023141}
W.-M. Huang and W.-M. Zhang,
Phys. Rev. Research \textbf{4}, 023141 (2022).

\bibitem{Yao24085114}
C.-Z. Yao and W.-M. Zhang,
Phys. Rev. B \textbf{110}, 085114 (2024).

\bibitem{Du25024139}
X. Du, J. Wang, and Y. Ma,
Phys. Rev. E \textbf{111}, 024139 (2025).

\bibitem{Hsi21085004}
J.-T. Hsiang and B.-L. Hu,
Phys. Rev. D \textbf{103}, 085004 (2021).

\bibitem{Mog22030335}
A. G. Moghaddam, K. P\"oyh\"onen, and T. Ojanen,
PRX Quantum \textbf{3}, 030335 (2022).

\bibitem{Bra1900444}
A. Brandenburger and K. Steverson,
Found. Phys. \textbf{49}, 444 (2019).

\bibitem{Fag260708655}
M. Fagotti,
arXiv:2607.08655 (2026).

\bibitem{Ste10}
S. J. Blundell and K. M. Blundell,
\textit{Concepts in Thermal Physics}, 2nd ed.
(Oxford University Press, Oxford, 2010).

\bibitem{Cam09392002}
M. Campisi, P. Talkner, and P. H\"anggi,
J. Phys. A: Math. Theor. \textbf{42}, 392002 (2009).

\bibitem{Tu08235311}
M. W.-Y. Tu and W.-M. Zhang,
Phys. Rev. B \textbf{78}, 235311 (2008).

\bibitem{Jin10083013}
J. S. Jin, M. W.-Y. Tu, W.-M. Zhang, and Y. J. Yan,
New J. Phys. \textbf{12}, 083013 (2010).

\bibitem{Zha12170402}
W.-M. Zhang, P.-Y. Lo, H.-N. Xiong, M. W.-Y. Tu, and F. Nori,
Phys. Rev. Lett. \textbf{109}, 170402 (2012).

\bibitem{SM}
See Supplemental Material at [URL will be inserted by publisher] for detailed
derivations of the TLS state-equivalent representations, the Lorentz--Drude
spectral representation and bound-state threshold, the numerical parameters
and consistency checks, and the HMF thermodynamics used in the main text.

\bibitem{Hua20165116}
Y.-W. Huang, P.-Y. Yang, and W.-M. Zhang,
Phys. Rev. B \textbf{102}, 165116 (2020).

\end{thebibliography}
\end{document}